# Brownian motion of nonlinear oscillator in van der Waals trap

Xiaofei Liu,* Fangyuan Chen, Zepu Kou, and Wanlin Guo

State Key Laboratory of Mechanics and Control of Mechanical Structures, Key Laboratory for Intelligent Nano Materials and Devices of Ministry of Education, Nanjing University of Aeronautics and Astronautics, Nanjing 210016, China

*E-mail: liuxiaofei@nuaa.edu.cn

**Abstract:** Van der Waals trap, a quantum fluctuation-induced potential characterized by short-range repulsive and long-range attractive forces, is intrinsically nonlinear. This work unveils the nonlinear effects on Brownian oscillators in the van der Waals trap using Langevin dynamics simulations and quasiharmonic approximations. While neither size- nor temperature-dependences of effective natural frequency is important for suspended plates of large areas, smaller ones with broader probability distributions are significantly softened and even a temperature-induced softening is observed. Despite the nonlinearity, the stiffness and the coefficient of friction are tunable by changing the thickness of coating and by modifying the size and the perforation condition of suspended plates, respectively, endowing the quantum trap with flexibilities of building up microscopic mechanical systems and probing near-boundary hydrodynamics.

## 1. Introduction

Van der Waals (vdW) or Casimir interaction, arising from the spatial gradient of zero-point electromagnetic energy [1,2], is important to a wide range of disciplines [3-8]. Two objects placed in vacuum are pulled towards each other, while those immersed in a dielectric medium other than vacuum can either attract or repel each other pending on the relative permittivities [9-11]. The force tends to decay with distance monotonously, irrespective of the attractive or repulsive nature. By combining the repulsion by a Teflon coating with the attraction by an underlying Au substrate, Zhao *et al.* [12] recently observes a nonmonotonous vdW potential exerted on a micro Au plate in ethanol environment. As the potential can confine the plate in a specific site along the perpendicular direction [12,13], it can be termed vdW trap. The quantum fluctuation-induced trap provides not only a stiffness but also an energy minimum, forming a mechanical oscillator without the attendance of a counteracting force of another type as required in the case of monotonous vdW potential [14,15].

As the quantum trap is formed within a liquid environment, the classical fluctuation is indispensable at a finite temperature [16,17]; namely, the oscillator is Brownian in nature. The "surface confinement" effect [18,19] and the backflow of disturbed liquid [20,21] also influence the frictional and random forces acting on the oscillator. Such hydrodynamic effects are usually probed using a dielectric sphere confined by the harmonic potential of optical tweezer [22,23]. The quantum fluctuation-induced trap offers a brand-new platform of probing the hydrodynamic effects, with advantages that probers of various shapes can be applied, the stiffness and the equilibrium separation can be rationally designed, and the trapping is passive without energy input.

In spite of the promising potential of building up microscopic mechanical systems and probing near-boundary hydrodynamic effects, the vdW trap is intrinsically nonlinear, with the force diverging to infinite at the close vicinity of the liquid-solid interface but vanishing at the infinite separation. At this point, the first priority of exploiting the coupling between the quantum fluctuation of electromagnetic field and the thermal fluctuation of fluid is to understand the nonlinear effect. In this Letter, we resort to Langevin dynamics (LD) simulations and quasiharmonic approximations to unveil the nonlinear effects on the Brownian motion of an Au plate immersed in ethanol and trapped by a Teflon-coated Au substrate. The effective natural frequency of oscillators beyond a critical size exhibits negligible size and temperature dependences. In stark contrast, smaller oscillators can be softened by an order of magnitude due to the broadened probability distributions and a temperature-induced softening is predicted by the statistical linearization and verified by the simulations. The tunability of the nonlinear Brownian oscillator under varying trap geometrics and hydrodynamic conditions is explored and the adequacy of the analytic approximations is elucidated.

## 2. Theories and models

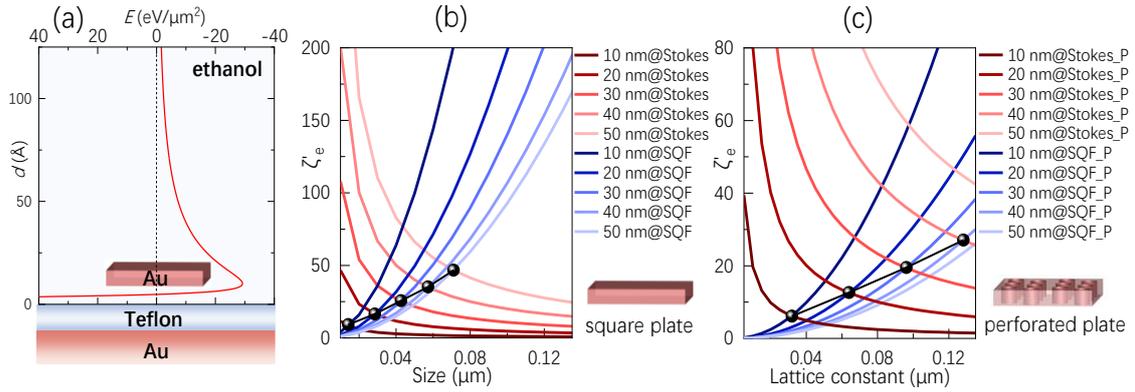

Fig. 1. Nonlinear Brownian oscillator in a vdW trap. (a) Configuration of vdW trap for Au plate immersed in ethanol. The displayed energy density-distance curve is for Au plate of 100 nm interacting with Teflon cover of 20 nm on Au substrate of 1 μm. (b) Effective damping ratio $\zeta_e$' at 293.15 K, as a function of square plate size and Teflon coating thickness. The red and blue lines are calculated using the Stokes law and the squeeze film damping, respectively. (c) $\zeta_e$' of perforated plate as a function of perforation lattice constant and Teflon coating thickness. The hole diameter-to-perforation lattice constant ratio is 0.8. The red and blue lines are calculated using the effective Stokes model and the effective squeeze film damping, respectively.

A typical energy density-distance relation of a vdW trap calculated with the Lifshitz theory [24,25] is illustrated in Fig. 1a. For an Au plate with a thickness of 100 nm interacting with a Teflon cover of 20 nm on an Au substrate of 1 μm, the energy minimum is 10.1 nm away from the liquid-solid interface, agreeing with Zhao et al.'s observation that the equilibrium separation, $h_e$, is about half the thickness of coating [12]. No obvious temperature ($T$) effect on $h_e$ (as has been found in heterogeneous multilayer configurations by Rodriguez et al. [26]) appears here; and the well depth changes by no more than 5% in the considered temperature range (Fig. S1a [27]). Since the temperature effect on potential is negligible in comparison with the temperature dependences of the effective natural frequency ($\Omega$), the potentials calculated at 293.15 K are exclusively used. All the energy density profiles applied are provided in Fig. S1b [27].

The stochastic motions along the perpendicular direction are simulated according to the Langevin equation

$$m\frac{dv(t)}{dt} = F(x) - \gamma v(t) + R(t), \quad (1)$$

where $m$, $v$, $\gamma$, $F(x)$ are mass, velocity, coefficient of friction, and restoring force, respectively. The Gaussian-type random force $R(t)$ is related to the friction via the classical fluctuation-dissipation theorem [28,29], $\sigma_R^2 = 2\gamma k_B T$, where $\sigma_R$ is the standard deviation of $R$ and $k_B$ is the Boltzmann constant. Positional power spectral densities (PSD) are derived from the last 0.2 ms temporal sequences in 0.3 ms simulations with a time step of $10^{-10}$ s. In order to mimic the stationary situation, each spectrum is an average over 60 runs.

Starting from the Fokker-Plank equation, a formulism equivalent to the Langevin equation, Nakajima and Zwanzig have proposed the projection operator method for stationary nonlinear Brownian oscillators [30,31], whose lowest and second lowest order approximants are the statistical linearization (or the independent quasiharmonic approximation, IQH) and the linearly coupled quasiharmonic approximation (LCQH) [32], respectively. The IQH gives a harmonic oscillator-like PSD

$$S_{IQH}(\omega) = \frac{\gamma \Omega^2 \langle x^2 \rangle_{st}}{m^2(\omega^2 - \Omega^2)^2 + \gamma^2 \omega^2}. \quad (2)$$

The effective natural frequency $\Omega$ is defined as $\Omega^2 = -\langle xF(x) \rangle_{st} / m \langle x^2 \rangle_{st}$, where the expectations are taken under the stationary distribution. Such a definition is equivalent to minimizing the mean-squared difference between the actual force and the effective linear spring force. The PSD by the LCQH taking account of the second derivative of potential reads

$$S_{LCQH}(\omega) = \frac{\gamma \Omega^2 \langle x^2 \rangle_{st}}{m^2[\omega^2 + \langle dF/dx \rangle_{st}]^2 + \gamma^2 \omega^2}. \quad (3)$$

The friction force by the fluid is considered by both the Stokes law, $\gamma_{Stokes} = 6\pi\eta R_e$, and the squeeze film damping (SQF) [33], $\gamma_{SQF} = 6\pi\eta R_e^4 / 4h_e^3$, where $R_e$ is the effective radius of a suspended square plate, and $\eta$ is kinetic viscosity of ethanol. The two models should be suitable for small and large plates, respectively, according the ratio of width to separation. The constant equilibrium separation $h_e$ is used here for the sake of a direct comparison between the simulation and the quasiharmonic approximations. To explore the damping regime that can be realized, Fig. 1b illustrates the effective damping ratio, $\zeta_e'=\gamma/2m\Omega'$, at a fixed thickness of plate ($t_{Au}$ =100 nm) but varying width ($w$) and coating thickness ($t_c$), calculated with $\eta$ at 293.15 K (1.2 cP). The prime symbol means that $\zeta_e$' is calculated with the converged $\Omega$' of a large enough plate, to solely include the effect of $\gamma$. Otherwise, $\Omega$ exhibits $w$ and $T$ dependences owing to the nonlinearity, unlike the natural frequency under a harmonic energy density that is independent of $w$ and $T$. $\zeta_e$' with $\gamma_{Stokes}$ is proportional to $w^{-1}$, while that with $\gamma_{SQF}$ is proportional to $w^2$. The intersection point identifies the critical width $w_c$, where an identical $\gamma_c$ is given by both models and $\zeta_e$' is the lowest at a specific $t_c$. Generally, the lowest $\zeta_e$' increases with $t_c$, from 9.2 at 10 nm to 35.2 at 40 nm.

Perforation, a strategy often applied to micro electromechanical systems [34], is considered to reduce the damping ratio (see Fig. S2 for the perforated plate with circular holes aligned within a hexagonal lattice [27]). To evaluate the friction of the perforated plate, an effective SQF model is used, $\gamma_{SQF\_P} = 3\sqrt{3}\eta w^2 a^2 C/4\pi h_e^3$, where the constant $C$ relies on the ratio of hole diameter $d$ to perforation lattice constant $a$, $\kappa = d/a$ [34]. As we considered a fixed $\kappa$ of 0.8, $C$ is 0.073. The resultant $\zeta_e$' is proportional to $a^2$. This leads to an unrealistically low $\zeta_e$' at extremely small $a$. To set a lower bound, an effective Stokes model assumes a proportional relation between friction and total hole edge length, $\gamma_{Stokes\_P} = 0.9067 \cdot 6\pi\eta R_e w\kappa/a$, with which $\zeta_e$' is proportional to $a^{-1}$. The intersection point of the two models gives the critical perforation lattice constant, $a_c$, upon which $\gamma_c$ and $\zeta_e$' are the lowest at a specific $t_c$. As illustrated in Fig. 1c for a perforated plate ($w$ = 0.5 μm, $t_{Au}$ = 100 nm), the lowest $\zeta_e$' at $t_c$ of 10 and 40 nm are reduced to 6.1 and 27.1, respectively. We point out that $\zeta_e$' could be further lowered by thinning the coating, searching for liquids of lower viscosities but dielectric functions similar to ethanol, or using a spherical Au bead.

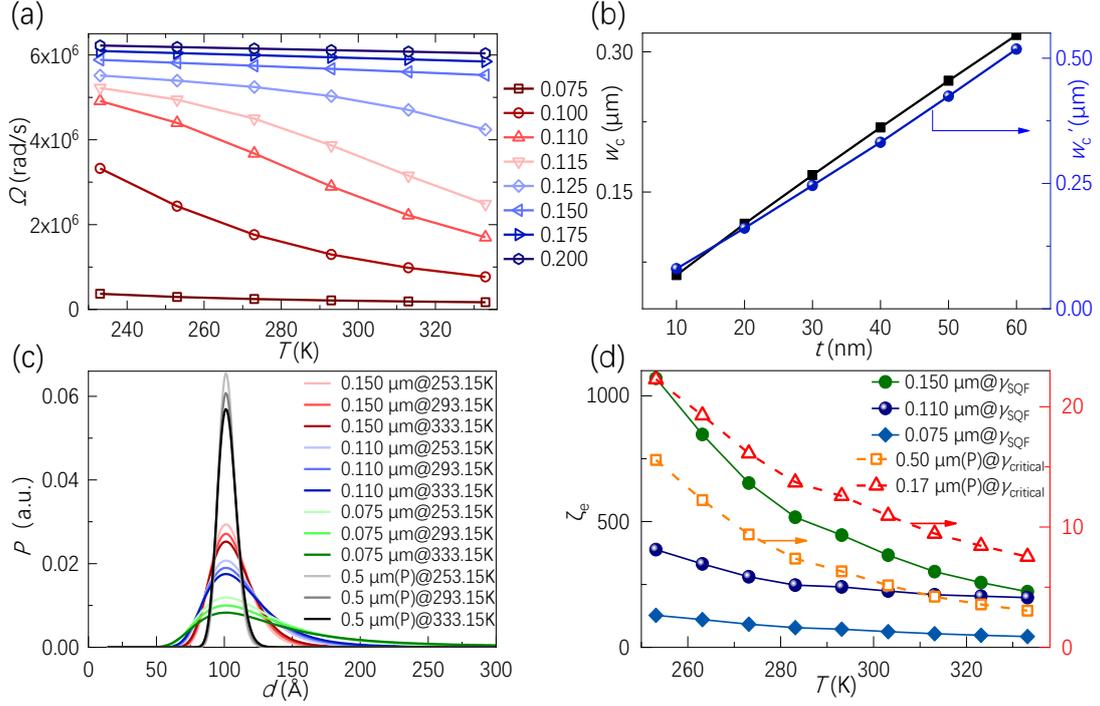

Fig. 2. Nonlinearity-induced size and temperature effects on the dynamical property. (a) Effective natural frequency as a function of temperature and width (in unit of μm), for Au slab of 100 nm trapped by Teflon coating of 20 nm. (b) The critical widths where $\Omega$ changes by twofold ($w_c$) and by no more than 5% ($w_c$') in the considered temperature range. (c) Probability distributions at different temperatures for square Au plate of varying width and perforated one of 0.5 μm with critical perforation lattice constant. (d) Effective damping ratio $\zeta_e$ as a function of temperature for Au plate of 100 nm with different widths. The perforated ones are of critical perforation lattice constants.

## 3. Results and discussions

Figure 2a presents the effective natural frequency of unperforated plates ($t_{Au}$ = 100 nm, $t_c$ = 20 nm) as a function of size and temperature. For a large enough plate, $\Omega$ of is insensitive to width. However, as the plate shrinks, $\Omega$ decreases dramatically from 6.22×10$^6$ rad/s at $w$ of 0.2 μm to 0.33×10$^6$ rad/s at $w$ of 0.075 μm at a fixed temperature of 253.15 K. Interestingly, plates of different sizes exhibit essentially different temperature dependences. For large ones, $\Omega$ is barely influenced by $T$. With $w$ of 0.2 μm, $\Omega$ changes by merely 2.9% in percentage and 0.18×10$^6$ rad/s in value in the considered temperature range (253.15 ~ 333.15 K). Reducing the size boosts the temperature effect. At a medium $w$ of 0.115 μm, $\Omega$ decreases by 52.5% in percentage and 2.74×10$^6$ rad/s in value. At the smallest $w$ of 0.075 μm, the absolute change is slight (0.20×10$^6$ rad/s), but the relative change is still important (54.0%). The size mentioned here should be understood relative to the thickness of coating, because both the critical widths at which $\Omega$ changes by twofold and by no more than 5% as illustrated in Fig. 2b are proportional to $t_c$. A fixed $t_c$ of 20 nm is adopted in the following for concise, but the basic physics should be general (Figs. S3a and S3b [27]).

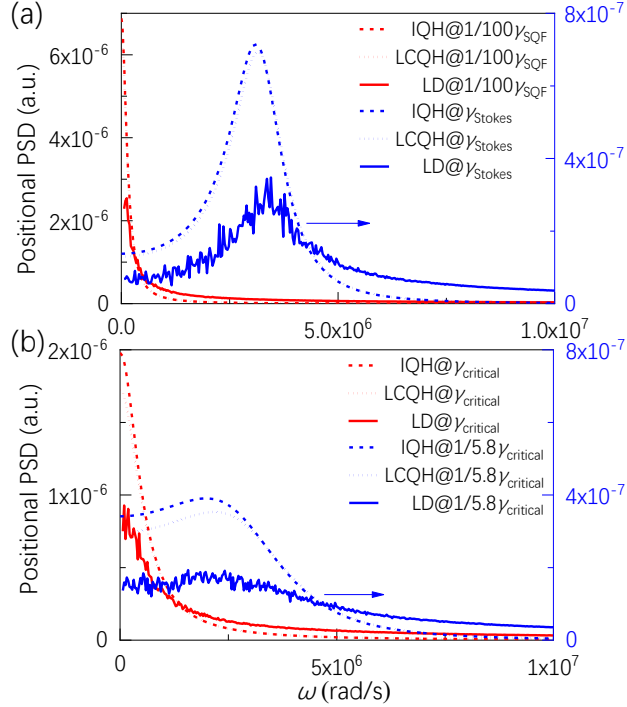

Fig. 3. (a) Positional power spectral density (normalized relative to the mean-square displacement) at 333.15 K for Au plate of 400 nm with width of 0.5 μm. The solid lines are results from the Langevin dynamics (LD) simulations. The dashed and dotted lines are results by the independent quasiharmonic (IQH) and linearly coupled quasiharmonic (LCQH) approximations, respectively. PSDs under $\gamma_{Stokes}$ and 1/100-fold $\gamma_{SQF}$ are presented. (b) Positional PSD at 333.15 K for the perforated Au plate with a critical perforation lattice constant. PSDs under $\gamma_c$ and 1/5.8-fold $\gamma_c$ are presented.

The size and temperature effects can be rationalized from the stationary probability distributions illustrated in Fig. 2c. For a large plate (the perforated one with $w$ of 0.5 μm), the probability at 253.15 K is narrowly distributed around the energy minimum with a full width at half maximum (FWHM) of 1.5 nm. Increasing temperature just slightly broadens the probability peak, which is the reason for the inertness of large plate. For narrower ones, the actual potential well depth is shallower and consequently the distributions are broadened. At $w$ of 0.075 μm, the FWHM at

253.15 K is enlarged to 6.4 nm and the mean position is shifted to the attractive side (Fig. S3c [27]). $\Omega$ is reduced consequently because of the fact that the stiffness at the attractive side is lower. Increasing temperature further broadens the distribution, which accounts for the sensitivity of $\Omega$ to $T$.

A trivial temperature effect that influences the dynamic behavior of the Brownian oscillator comes from the $T$-dependent viscosity. Figure 2d illustrates the effective damping ratio $\zeta_e$, taking account of not only the size and temperature effects on $\Omega$ but also the $T$-dependent $\eta$. $\zeta_e$ of large plates (unperforated one with $w$ of 0.15 μm and perforated one with $w$ of 0.5 μm, $t_{Au}$ = 100 nm) increases with reducing $T$, with a scaling trend complying with the $\eta$-$T$ relation (Fig. S4 [27]). For smaller ones (unperforated ones with $w$ of 0.11 or 0.075 μm and perforated one with $w$ of 0.17 μm), the temperature effect on $\Omega$ somewhat retards the increase of $\zeta_e$ with reducing $T$.

To look into the nonlinear effect on the positional spectrum, the simulated and analytic PSDs of a large plate ($w$ = 0.5 μm, $t_{Au}$ = 400 nm) at 333.15 K with $\eta$ of 0.58 cP are presented in Fig. 3a. To represent both the underdamped and overdamped regimes, two coefficients of friction $\gamma_{Stokes}$ and 1/100-fold $\gamma_{SQF}$ are considered ($\gamma_{SQF}$ is so large that a huge number of steps is required for a reliable sampling). The simulated PSD under $\gamma_{Stokes}$ exhibits a resonant peak at ~3.4×10$^6$ rad/s (close to $\Omega$), while that under 1/100-fold $\gamma_{SQF}$ decays with frequency ($\omega$). The simulation results are consistent with the quasiharmonic approximations, since $\zeta_e$' of the two cases are evaluated to be 0.22 and 11.98, respectively. Indeed, the IQH and LCQH spectra qualitatively agree with the simulated ones, as highlighted by the unambiguously reproduced resonant peak under $\gamma_{Stokes}$. Quantitatively, the low-frequency densities ($\omega$ < 4.3×10$^6$ and < 0.35×10$^6$ rad/s in the two cases, respectively) are overestimated by the approximations, whereas the high-frequency ones are underestimated. The quantitative deviation stems from the fact that the stiffness is not a constant but a function of displacement. Figure 3b presents the PSDs of a plate perforated with $a_c$ ($w$ = 0.5 μm, $t_{Au}$ = 400 nm), simulated under $\gamma_c$ and 1/5.8-fold $\gamma_c$. The simulated spectra qualitatively agree with the analytic ones characterized by $\zeta_e$' of 3.05 and 0.53, respectively.

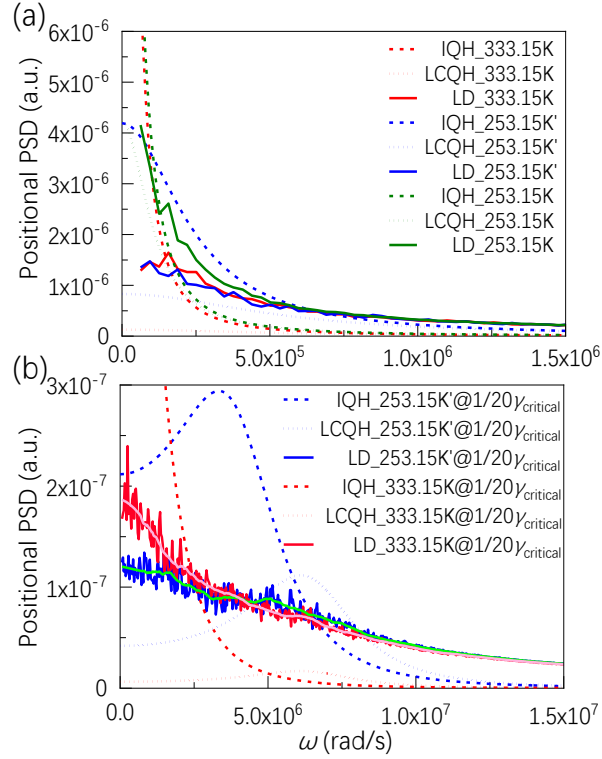

Fig. 4. (a) Positional PSD at 253.15 and 333.15 K for the perforated Au plate of 0.17 μm with a critical perforation lattice constant. The prime symbol denotes that the spectrum at 253.15 K is calculated with the same viscosity at 333.15 K (0.58 cP). (b) Positional PSD of the same perforated Au plate under 1/20-fold $\gamma_c$. The pink and green lines are the smoothed LD spectra.

Figure 4a shows the positional PSDs of a small plate (perforated one with $w$ of 0.17 μm and $a_c$, $t_{Au}$ = 100 nm). To distinguish the nonlinearity-induced temperature effect from the trivial temperature dependence, the PSDs at 333.15 and 253.15 K are simulated with the same $\eta$ of 0.58 cP. The low-frequency SPDs are higher at 333.15 K, indicating a temperature-induced softening. The IQH spectrum at 253.15 K evaluated with $\eta$ of 0.58 cP qualitatively agrees with the simulated one, except for the overestimation and underestimation of the low- and high-frequency densities as appear in the case of large plate. Nevertheless, it must be noted that the IQH overestimates the low-frequency PSDs at 333.15 K more severely, meaning that the temperature-induced softening can be overestimated by the statistical linearization. Considering the actual $\eta$ at 253.15 K (2.95 cP), the simulated low-frequency PSDs become higher than those at 333.15 K with $\eta$ of 0.58 cP, which makes sense as $\zeta_e$ are 39.29 and 22.33 in the two cases, respectively.

To evaluate the nonlinear effects on smaller plates at underdamped regime, Fig. 4b presents the spectra of the same perforated plate but under 1/20-fold $\gamma_c$. At 253.15 K with $\eta$ of 0.58 cP, a resonant peak at ~$5.1\times10^6$ rad/s is clearly observed, although the spectrum deviates from the harmonic one with $\zeta_e$ of 0.386 or with $\zeta_e$' of 0.305. The resonant frequency is lower than the converged $\Omega$ of large plate ($6.22\times10^6$ rad/s), verifying the small size-induced softening. However, the resonant frequency is higher than the estimated $\Omega$ ($4.40\times10^6$ rad/s), again indicating that the statistical linearization overestimates the softening. At 333.15 K with $\eta$ of 0.58 cP, the simulated PSD decays with $\omega$, with the low-frequency densities being higher than those at 253.15 K. Especially, no resonant peak near the estimated $\Omega$ ($2.22\times10^6$ rad/s) is observed. These results verify the temperature-induced softening at underdamped regime. Beyond our expectation, the IQH fails to predict the correct spectra even qualitatively. The IQH spectrum at 253.15 K gives an incorrect trend of low-frequency density and an inaccurate resonant frequency. The low-frequency IQH density ($6.1\times10^5$ rad/s) at 333.15 K can even be five-fold higher than the simulated value. On the other hand, the LCQH spectra for smaller oscillators completely diverges from the IQH spectra, in contrast to the situation for large plate where the two approximations give consistent results. With $\gamma_c$, the LCQH underestimates the SPDs over the whole frequency range. With 1/20-fold $\gamma_c$, it underestimates the SPDs at 333.15 K over the whole range, but kind of reproduces the resonant peak at 253.15 K.

## 4. Conclusion

To conclude, the work unveils the nonlinearity-induced size and temperature effects on the dynamical properties of oscillators confined in quantum fluctuation-induced vdW traps and subjected to classical thermal fluctuations. While the nonlinearity results in quantitative deviations of positional spectra from the quasiharmonic approximations for large suspended plates, it leads to both size-induced and temperature-induced softening of smaller oscillators, due to the broadened probability distributions. Comparison between the simulated and analytic spectra elucidates that the quasiharmonic approximations qualitatively work for large oscillators but fail for smaller ones with strong nonlinear effects. Our results shed a light on the promising potential of the quantum trap in designing of microscopic mechanical systems and probing of near-boundary hydrodynamic effects. Accurate analytic solution of the intrinsic nonlinear system and rich physics as related to the memory effect of fluid deserve future efforts.

**Acknowledgments:** The work is supported by the National Natural Science Foundation of China

(Grants No. 12072152 and No. 11702132), the National Key Research and Development Program of China (Grant No. 2019YFA0705400).

# Appeddix A: Supplementary Material for "Brownian motion of nonlinear oscillator in van der Waals trap"

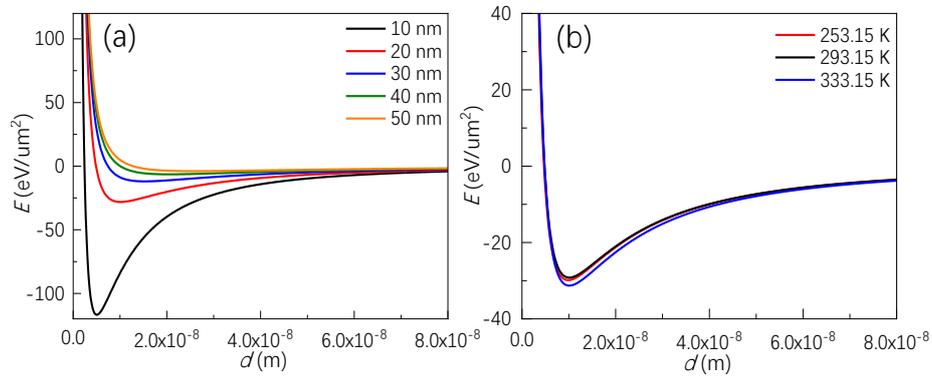

Fig. S1. (a) Van der Waals energy density as a function of separation at room temperature under varying Teflon coating thickness. (b) Van der Waals energy density as a function of separation under varying temperatures for a fixed Teflon coating of 20 nm. For both (a) and (b), the thicknesses of the suspended Au plate and the Au substrate are 100 nm and 1 μm, respectively. The van der Waals energies are calculated by the Lifshitz theory [1,2], using dielectric functions of Teflon, ethanol, and Au [3-5].

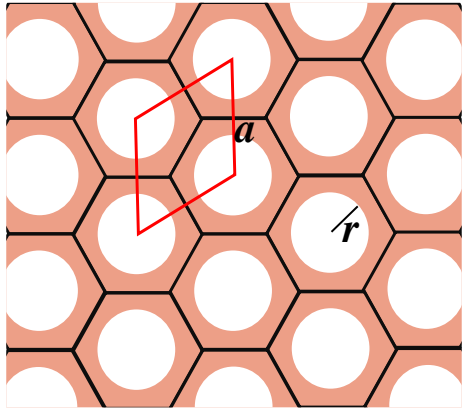

Fig. S2. Illustration of square Au plate perforated with hexagonal lattice of circular hole.

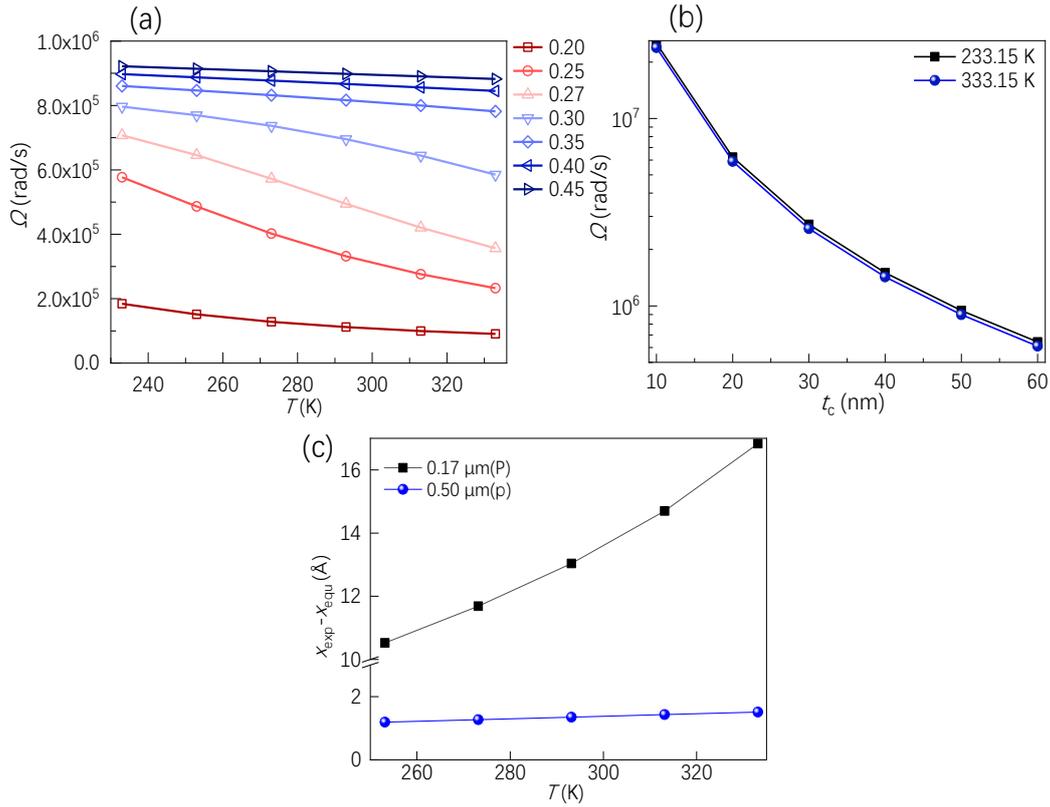

Fig. S3. (a) Effective natural frequency $\Omega$ as a function of temperature for a square Au plate of 100 nm with varying width, trapped by Teflon coating of 50 nm. (b) The converged $\Omega$ of large enough plate as a function of Teflon coating thickness. The converged $\Omega$ is estimated for the Au plate of critical width, with which $\Omega$ changes by no more than 5% at 233.15 and 333.15 K. (c) The expected displacement relative to the equilibrium position as a function of temperature, for the large and small perforated plates (the widths are 0.50 and 0.17 μm, respectively, and the Teflon coating thickness is 20 nm).

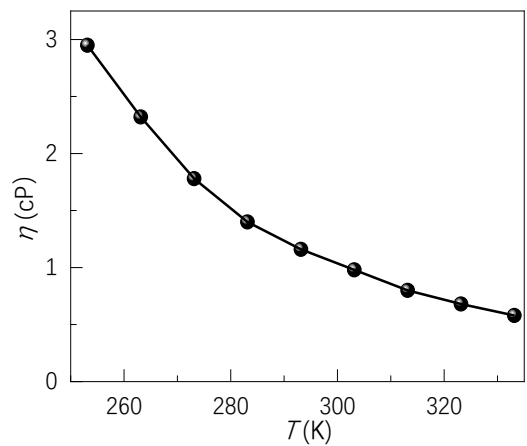

Fig. S4. Kinetic viscosity of ethanol at varying temperature. The data are obtained from the reference

[6].

## References of the supplementary material